\documentclass[%
 reprint,
 amsmath,amssymb,
 aps,
]{revtex4-2}

\usepackage[nottoc]{tocbibind}
\usepackage[english]{babel}
\usepackage[utf8]{inputenc}
\usepackage{blindtext}
\usepackage[T1]{fontenc}
\usepackage[utf8]{inputenc}
\usepackage{cool}
\usepackage{titlesec}
\usepackage{amsmath}
\usepackage{physics}
\usepackage{mathtools}
\usepackage{graphicx}
\usepackage{caption}
\usepackage{amsmath}
\usepackage{float}
\usepackage{graphicx} 
\usepackage{color}
\usepackage{yfonts}
\usepackage{dcolumn}
\usepackage{bm}

\begin{document}


\title{Application of Quantum Graph Theory to Metamaterial Design: \\
Negative Refraction of Acoustic Waveguide Modes} 

\author{T. M. Lawrie$^1$, T. A. Starkey$^2$, G. Tanner$^1$,  D. B. Moore$^2$, P. Savage$^2$, G. J. Chaplain$^2$,\\
{$^{1}$School of Mathematical Sciences, University of Nottingham, United Kingdom}\\
{$^{2}$Centre for Metamaterial Research and Innovation, Department of Physics and Astronomy, University of Exeter, United Kingdom}}

\begin{abstract}

We leverage quantum graph theory to quickly and accurately characterise acoustic metamaterials comprising networks of interconnected pipes. Anisotropic bond lengths are incorporated in the model that correspond to space-coiled acoustic structures to exhibit dispersion spectra reminiscent of hyperbolic metamaterials. We construct two metasurfaces with embedded graph structure and, motivated by the graph theory, infer and fine-tune their dispersive properties to engineer non-resonant negative refraction of acoustic surface waves at their interface. Agreement between the graph model, full wave simulations, and experiments bolsters quantum graph theory as a new paradigm for metamaterial design.

\end{abstract}

\maketitle

\section{Introduction}

Metamaterials have proven to provide a unique tool for the tailored control of wave propagation. These materials are typically composed of subwavelength building blocks (``meta-atoms"), often periodically repeated in space, which can have exotic effective material properties. Extensive research efforts are ongoing to optimise geometries and parameter regimes for the design of the constituent building blocks. Recently in acoustics, metasurfaces for the control of sub-sonic sound have received attention \cite{Beadle2018,Beadle2019,Ward2019,Cselyuszka2019,Moore2023, Moore2024}, based on the canonical system of coupled resonant cavities embedded in a fluid with Neumann (sound-hard) boundary conditions on the walls \cite{Kelders1998}. These systems can support localised acoustic surface waves (ASWs) that propagate along the interface between the metasurface and the surrounding fluid, when cavities are coupled via near field coupling, or Acoustic waveguide modes (AWMs) when resonators are coupled via waveguides.

There is much interest in fast and accurate design methods for metamaterial constituents. A relatively recent technique introduced to this field is a quantum graph formulation of metamaterial design \cite{lawrie2022quantum}, related but distinct to the transfer matrix method that is frequently utilised in acoustic metamaterial modelling \cite{jimenez2017rainbow,jimenez2021transfer}. In this letter, we highlight the efficacy of the quantum graph model in this context. Quantum graph theory, defined as metric graphs endowed with differential operators, have a long history in mathematics, physics, and both theoretical chemistry and biology \cite{Pauling36, RS53, Coulson54, Montroll70, Roth83, Alexander83, Below88}. For this work we consider the scattering formulation first introduces in 1997 \cite{KS97}, which serves as a powerful tool for studying quantum and wave phenomena. The theory describes the wave properties of networks constructed from edges, endowed with the Schr\"odinger operator, connected by vertices that serve as scattering sites, given specific boundary conditions \cite{BK13}. A key advantage to this discrete scattering language, is that the eigenvalue condition can be written in terms of a secular equation involving the determinant of a unitary matrix of dimension equal to twice the number of edges on the graph. Similarly, the scattering matrix of an open quantum graph can be given in terms of a closed-form expression involving matrices of dimension equal to the number of open channels \cite{KS97, BK13}. This simplicity allows one to express the scattering properties and even the Green’s function in a closed semi-analytic form \cite{lawrie2023closed, BG01}. Quantum graph models have found a great number of applications including the study of spectral statistics \cite{kottos1999periodic}, quantum chaos \cite{GS06}, modelling the vibrations of coupled plates \cite{BCT18}, formulating quantum random walks \cite{Kempe03, Tan06}, quantum search algorithms \cite{HT09}, nerve impulse transmission \cite{nicaise1985some} and model various protein \cite{gadiyaram2019quantum, kannan1999identification} and molecular structures \cite{griffith1953free, exner2001bound}. Indeed the model has been applied to both finite and infinite periodic systems, including ladder \cite{delourme2017trapped}, square \cite{hein2009wave}, hexagonal \cite{ figotin1998spectral, axmann1999asymptotic, amovilli2004electronic}, and Cairo lattices \cite{baradaran2023cairo}. With natural applications in photonic \cite{kuchment2002differential} and phononic crystals \cite{figotin1996band}. Our aim in this work is to extend and cement the usefulness of quantum graph methods in the context of metamaterial design.

We consider an infinite periodic quantum graph endowed with the Helmholtz equation \cite{lawrie2022quantum,lawrie2023engineering} that can naturally be mapped to describe the propagation of acoustic waves in a closed system of connected cylindrical wave\-guides (below the first cut-off frequency) \cite{coutant2023topologically}. We vary the topology of the underlying graph by including anisotropy that is, by changing the edge lengths in one of the two space dimensions. This is achieved in the physical system by using embedded space-coiled waveguides \cite{Liang2012}, a design conventionally used to facilitate absorption. Here, we shall instead utilise it to manipulate sound propagation confined to a surface. 

The notion of embedding the underlying structure has benefits in terms of fabrication and enhanced sensor response \cite{Moore2023,Jankovic2021,choudhury2021}. Theory, simulation (based on quantum graph models and finite element analysis), and experiment are presented of the resulting acoustic metasurface. We design and achieve non-resonant negative refraction \cite{luo2002all} at an interface between two metasurfaces by breaking the phase symmetry of the unit cell through variation of the underlying graph edge lengths. The result is an acoustic metasurface with an embedded network of graph-like connections that emulates a hyperbolic metamaterial \cite{smith2003electromagnetic}.

\section{Quantum Graph Formulation}

\begin{figure}[h!]
\centering
\includegraphics[width = 0.475\textwidth]{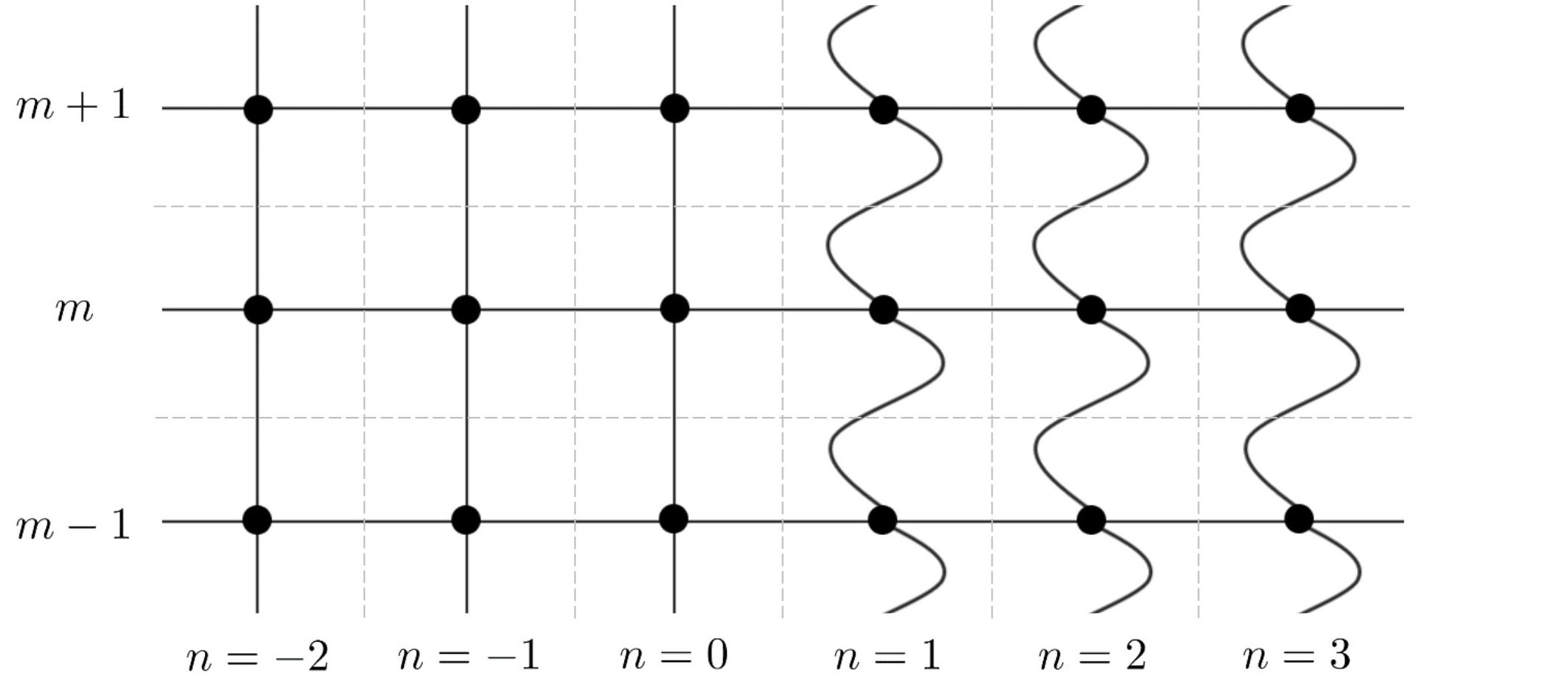}
\caption{Graph representation of two metamaterials connected at the interface between $n = 0$ and $n=1$. Note that the edge lengths in material on the RHS are different in the two space directions.}
\label{fig:Graph Setup}
\end{figure}
\begin{figure*}
    \centering
    \includegraphics[width = 0.95\textwidth]{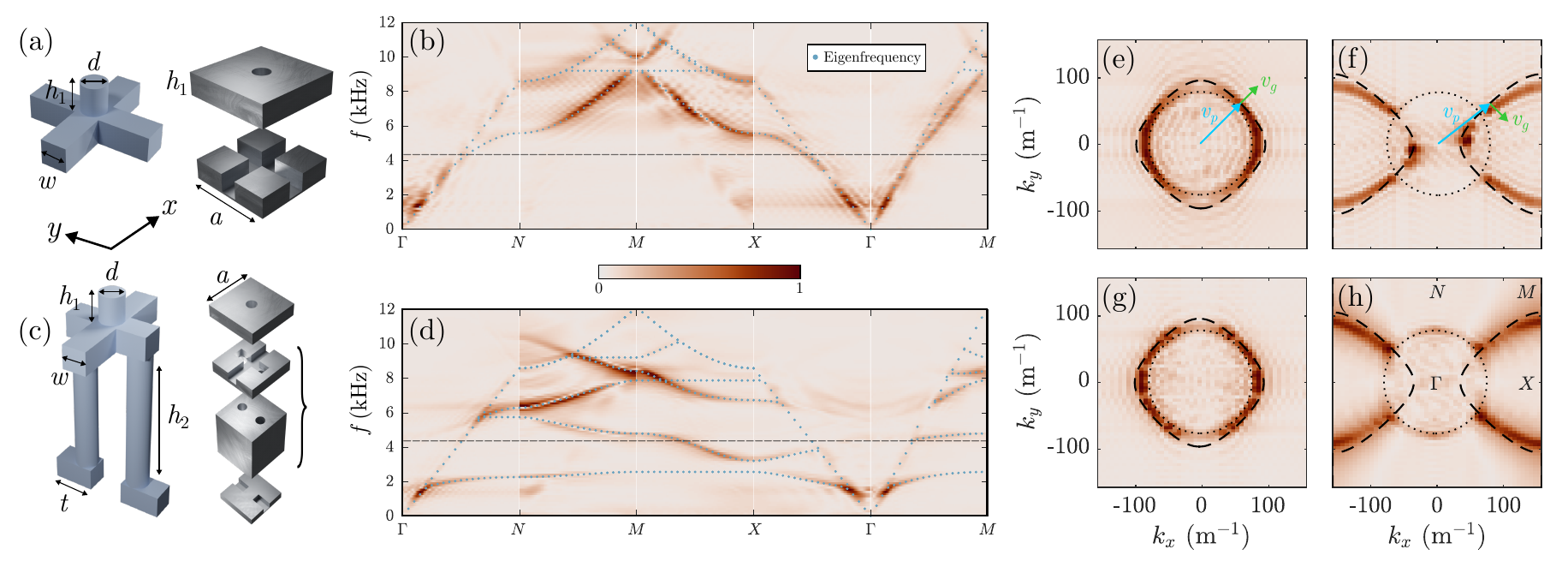}
    \caption{(a) Schematic of the pipe network as used in the experiment consisting of the fluid path (blue regions) and milled plates. The network is formed of pipes with a square cross-section of width $w = 4$ mm. The radius of the cavity connecting the network to the surface has diameter $d=2$ mm and height $h_1 = 5$ mm. The unit cell pitch is $a = 20$ mm. (b) Experimental Fourier spectra, normalised per wavenumber along each path in the IBZ. Overlaid are FE solutions (dotted line) below the sound cone. Note that the axis is rotated to be consistent with the experimental stage (i.e. $X \leftrightarrow N$). (c) Anisotropic counterpart to (a), with additional dimensions $h_2 =12.5$ mm and $t = 7$ mm. The exploded view of the plate shows the combination of milling/drilling (plates highlighted by curly brackets are one piece). (d) Corresponding Fourier spectra to (c). (e,f) Numerical isofrequency contours obtained at $4.2$ kHz (dashed line in (b,d)) for the configurations in (a,c) respectively. (g,h) Experimental counterparts - dotted circles show the sound cone and dashed lines show the analogous contours from the scaled graph model (closed system).} 
    \label{fig:exp}
\end{figure*}

In this section, we introduce the quantum graph formulation that models wave scattering in a periodic arrangement of thin tubes as a model for a metamaterial, see \cite{lawrie2022quantum} for further details. Quantum graph theory can be viewed as the limiting case of networks formed of channels of width $\epsilon$ where the differential operator on the manifold converges to the one-dimensional graph operator in the thin channel limit $\epsilon \rightarrow 0$ \cite{kuchment2001convergence,coutant2020robustness,coutant2021acoustic}. This theory will remain accurate for sufficiently thin tubes or indeed in the long wavelength regime. We shall first compare this theory to a network of enclosed sound hard channels, or pipes, filled with air. We then use this model to infer the behaviour of a network of pipes connected to free space , forming a metasurface (e.g. Figs.~\ref{fig:exp}\&~\ref{fig:SimExp1}); in this instance the theory does not fully apply, as (at present) the graph is not defined in the continuum. Despite this, we see that the metasurface is endowed with the dispersive properties predicted by the graph and use it as a predictive tool to model the wave dispersion of coupled waveguide modes.

To construct the illustrated quantum graph in Fig.\ref{fig:Graph Setup} we consider the eigen solutions of both the left and right hand sides as independent infinite periodic meshes which we later combine following the methodology in \cite{lawrie2022quantum}. For this, consider an infinite square periodic arrangement of vertices embedded in a 2D Euclidean space with period $\ell$. The space has respective horizontal and vertical continuous coordinates $x$ and $y$, while the vertices have respective horizontal and vertical discrete coordinates $n$ and $m$. Each vertex is connected horizontally and vertically by finite edges to the left($l$), right($r$), down($d$) and up($u$) of the vertex. Let us define the set of edges attached to a given vertex as $\mathcal{S} = \{l,r,d,u\}$ with valency (the number of connected edges at a given vertex) $v$, with $v = |\mathcal{S}|$; here we have $v=4$. The edges that form the vertex star are assigned a metric $\mathcal{L} = \{\ell_x,\ell_x,\ell_y,\ell_y\}$. The metric does not have to be equal to the vertex period $\ell$, allowing for space coiled channels; this is illustrated in Fig.\ref{fig:Graph Setup} for vertical edges ($d$ and $u$) for $n \geq 1$. For each edge $e\in\mathcal{S}$, we introduce an edge coordinate that exists within the domains $z_{nm,l} \equiv z_{nm,r} \in [0,\ell_x]$ and $z_{nm,d} \equiv z_{nm,u} \in [0,\ell_y]$, with $z_{nm,e} = 0$ at the vertex. The metric graph is turned into a quantum graph by the addition of a self-adjoint differential operator together with a set of boundary conditions on the graph vertices. The self-adjointness in physical terms ensures conservation of current at the vertex. For this work we consider the operator as the negative 1D Laplacian, representing some arbitrary scalar field. This choice makes the formulation applicable to a broad class of problems, such as in acoustics, optics or indeed single particle free space quantum mechanics. The self-adjointness implies unitary time evolution $e^{-i\omega t}$ where $\omega$ is some positive real number that represents frequency. Through a separation of variables the allowed space of functions on the graph edges must satisfy the Helmholtz wave equation, that is,
\begin{equation}\label{Helmholtz}
\left(\frac{\partial^{2}}{\partial z_{nm,e}^{2}} + k^{2}\right)\psi_{nm,e}(z_{nm,e}) = 0,
\end{equation}
where $k$ represents the wave number. The relationship between $k$ and $\omega$ depends on the physical system being modelled. For quantum mechanics the time evolution is given by the operator $i\partial / \partial t$ with free space quadratic dispersion $k^2 \propto \omega$. While in classical mechanics the time evolution is given by the operator $\partial^{2} / \partial t^{2}$ with linear free space dispersion $k \propto \omega$, with the constant of proportionality being the wave speed on the graph edge.
Here $\psi_{nm,e}$ is the wave function at vertex $nm$ on edge $e$, with solution given as a superposition of counter-propagating plane waves heading ``in'' or ``out'' of the vertex, that is,
\begin{equation}\label{Helmholtz Solution}
\psi_{nm,e}(z_{nm,e}) = \text{e}^{i\boldsymbol{k}\cdot\boldsymbol{r}}\left(a_{e}^{\text{out}}\text{e}^{ikz_{nm,e}} + a_{e}^{\text{in}}\text{e}^{-ikz_{nm,e}}\right).
\end{equation}
Here, $a_{e}^{\text{out/in}}$ represents a complex wave amplitude on edge $e$ and $\text{e}^{i\boldsymbol{k}\cdot\boldsymbol{r}} = \text{e}^{i(k_{x}n + k_{y}m)\ell}$ represents the Bloch phase of the periodic graph, with horizontal and vertical quasi momentum $k_x$ and $k_y$ \cite{kittel2005introduction}. While one is free to define any vertex boundary conditions, provided self-adjointness is maintained, (see \cite{kostrykin1999kirchhoff} for the most general matching conditions), we consider for this work Kirchhoff-Neumann (KN) boundary conditions. Explicitly, we require:
\begin{enumerate}
    \item \textit{The wave functions are continuous at the vertex}
        \begin{equation}\label{vertex continuity}
        \psi_{nm,e}(0) = \psi_{nm,e'}(0) .
        \end{equation}
   \item \textit{The outgoing derivative of the function on each edge $e$ at the vertex must satisfy},
    \begin{equation}\label{vertex gradient}
     \sum_{e \in \mathcal{S}} \frac{\partial \psi_{nm,e}}{\partial z_{nm,e}}(0) =
        0 .\\
    \end{equation}
\end{enumerate}
By substituting Eq.(\ref{Helmholtz Solution}) into (\ref{vertex continuity}) and (\ref{vertex gradient}), we express the vector of outgoing wave amplitudes $\boldsymbol{a}^{\text{out}} = \left(a_{l}^{\text{out}},a_{r}^{\text{out}},a_{d}^{\text{out}},a_{u}^{\text{out}}\right)^T$ in terms of the vector of incoming wave amplitudes $\boldsymbol{a}^{\text{in}} = \left(a_{l}^{\text{in}},a_{r}^{\text{in}},a_{d}^{\text{in}},a_{u}^{\text{in}}\right)^T$ at the vertex, via the scattering matrix $\hat{S}$, which performs the mapping,
\begin{equation}\label{scattering}
    \boldsymbol{a}^{\text{out}}
    =\hat{S}
    \boldsymbol{a}^{\text{in}}.
\end{equation}
Here the $pq^{\text{th}}$ elements of the vertex scattering matrix are,
\begin{equation}\label{Neumann Boundary Conditions}
    \hat{S}_{pq} = \frac{2}{{v}} - \delta_{pq}.
\end{equation}
As for the entire structure, we can evaluate the solution at a given vertex in terms of the solutions at its neighbouring vertices, explicitly, 
\begin{equation} \label{Edge and Bloch Phase}
\begin{split}
\psi_{nm,l}(\ell_x) &= \text{e}^{-ik_x\ell}\psi_{nm,r}(0) \\
\psi_{nm,r}(\ell_x) &= \text{e}^{ik_x\ell}\psi_{nm,l}(0) \\
\psi_{nm,d}(\ell_y) &= \text{e}^{-ik_y\ell}\psi_{nm,u}(0) \\
\psi_{nm,u}(\ell_y) &= \text{e}^{ik_y\ell}\psi_{nm,d}(0). \\
\end{split}
\end{equation}
This allows one to map the outgoing wave amplitudes to the incoming wave amplitudes as, 
\begin{equation}\label{Bloch conditions}
\boldsymbol{a}^{\text{in}}
=
\hat{B}(k,k_x,k_y)
\boldsymbol{a}^{\text{out}},
\end{equation}
where
\begin{equation}\label{Bloch condition definition 1}
\hat{B} = 
\begin{pmatrix}
    0 & \text{e}^{i(k\ell_x-k_x\ell)} & 0 & 0 \\
    \text{e}^{i(k\ell_x+k_x\ell)} & 0 & 0 & 0 \\
    0 & 0 & 0 & \text{e}^{i(k\ell_y-k_y\ell)} \\
    0 & 0 & \text{e}^{i(k\ell_y+k_y\ell)} & 0 \\
\end{pmatrix}.
\end{equation}
By substituting \eqref{scattering} into \eqref{Bloch conditions}, we get an eigenvalue condition in terms of the quantum map $\hat{U} = \hat{B}\hat{S}$, that is,
\begin{equation}\label{Eigen function}
    \left[\hat{\mathbb{I}} - \hat{U}(k,k_x,k_y)\right]
    \boldsymbol{a}^{\text{in}}
    = \boldsymbol{0}.
\end{equation}
Dispersion curves can be obtained from the secular equation,
\begin{equation} \label{eqn:secular}
    \text{det}\left[\hat{\mathbb{I}} - \hat{U}(k,k_x,k_y)\right] = 0.
\end{equation}
Solving for the above roots with $\hat{S}$ given by Eq.\ (\ref{Neumann Boundary Conditions}) leads to the condition
\begin{equation}\label{Graph Dissipative Properties}
    \text{sin}(k(\ell_x + \ell_y)) = \text{sin}(k\ell_y)\text{cos}(k_x\ell) + \text{sin}(k\ell_x)\text{cos}(k_y\ell),
\end{equation}
which for the isotropic case, $\ell_x = \ell_y = \ell$, leads to the `free-space' solutions
\begin{equation}\label{Graph Free Space Dissipative Properties}
    2\text{cos}(k\ell) = \text{cos}(k_x\ell) + \text{cos}(k_y\ell).
\end{equation}
\begin{figure*}
\centering
\includegraphics[width = 0.95\textwidth]{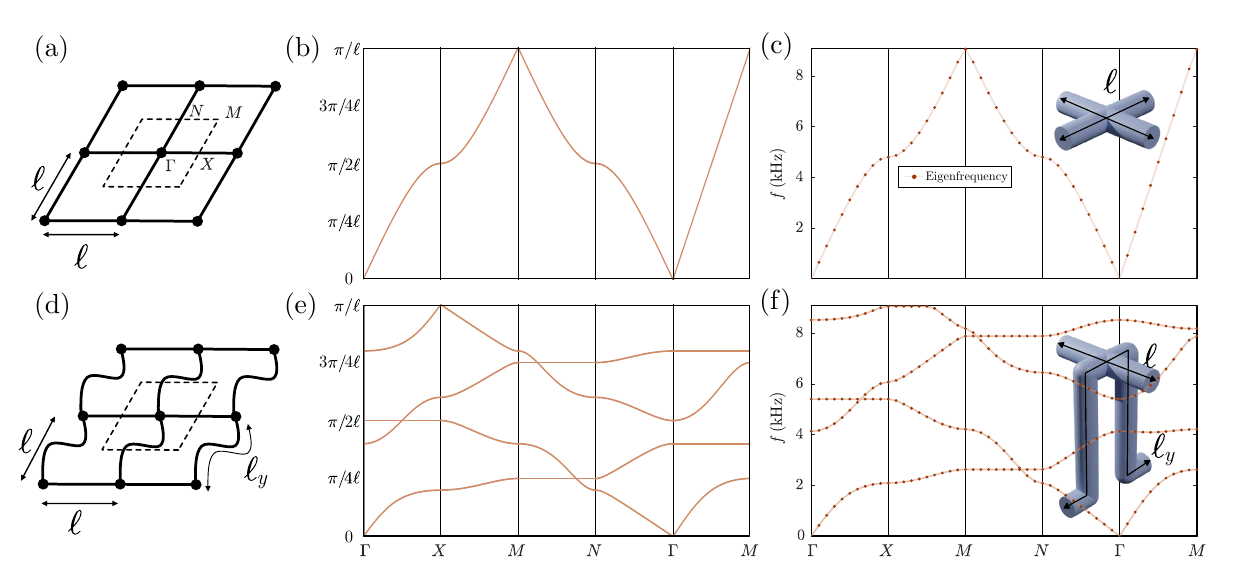}
\caption{\label{fig:Solution Space} Comparisons of graph model and finite element dispersion curves. (a) A section of the infinite isotropic graph structure ($\ell_x = \ell_y = \ell$). Vertices connect to nearest neighbours in a mesh of edges of length $\ell$. The unit cell is shown as a rectangle around the central vertex. (b) Corresponding 
dispersion curves along the Irreducible Brillouin Zone (IBZ) (directions marked in (a)). (c) Corresponding FEM dispersion curves for acoustic waves in a pipe network with pipes of radius $2$ mm and $\ell_x = \ell_y = \ell = 20$ mm; the unit cell is shown in the inset. (d) Anisotropic graph counterpart to (a), such that $\ell_x = \ell$, $\ell_y = 4\ell$. (e) Corresponding dispersion curves. (f) Anisotropic counterpart to (c), with coiled path in $y$ such that $\ell_y = 4\ell$.}
\label{fig:disps}
\end{figure*}
The dispersion curves \eqref{Graph Dissipative Properties} are shown in Fig.~\ref{fig:disps}(b) and (e) for both isotropic and anisotropic materials with generalised non-dimensional material parameters. In the anisotropic case, we chose $\ell_x =\ell$, $\ell_y = 4 \ell$. 

The above analytic solutions of the quantum graph model can be related to wave solutions of quasi 1D scalar fields. This is demonstrated by comparing with numerical solutions obtained via the finite element method (FEM) using COMSOL Multiphysics \cite{Comsol}, see Fig.~\ref{fig:disps}(c) and (f). Here, we set up an equivalent physical system of connected acoustic pipes in both the isotropic and anisotropic configuration. An edge in the graph model now corresponds to a pipe (of fixed radius) and a vertex corresponds to an intersection of pipes. The change in impedance at the connecting region serves as the scattering site which qualitatively agrees with the KN boundary condition \eqref{Neumann Boundary Conditions} in the graph model, providing we operate below the first cut-off frequency of the pipes. The graph structures accurately model the acoustic wave propagation in analogous pipe networks solved via FEM, as can be seen from comparing Fig.~\ref{fig:disps}(b) and (e) with Fig.~\ref{fig:disps}(c) and (f). Deviations start to occur with increasing frequency as expected, as we neglect spatial inhomogeneities such as bends in the graph model and other finite size effects \cite{xie2013tapered}. This can be seen when comparing the upper bands in Fig.~\ref{fig:disps}(c) and (f). The graph model (in this form) is insensitive to the geometry of the anisotropic path, as it is parameterised only in terms of it's length, $\ell_{y}$. Sharp bends can however be included in the graph model and it can, of course, be recast to correspond directly to the acoustic pressure field $p$ i.e. $\psi \mapsto p$ such that $k = \omega/c$ where $\omega$ is radian frequency and $c$ the speed of sound. Entries in the scattering matrix are then scaled with the characteristic acoustic impedance to relate the pressure and velocity wave amplitudes in the time-harmonic regime \cite{jimenez2021transfer}. Given the intricacies that arise in the experimental measurement (see Section~\ref{sec:exp}) we proceed with the generalised case. 

To see the effect of negative refraction, we couple the two materials at a common interface along the $y$-direction as shown in Fig.~\ref{fig:Graph Setup} with results presented in the next section. In the graph case, the coupling conditions can be given analytically by satisfying an equivalence condition between the eigenstates of the different lattices on the edges across the interface \cite{lawrie2022quantum}.

\section{Experimental validation}
\label{sec:exp}

\subsection{Characterising the metasurfaces}

Experimentally characterising a mesh of closed pipes such as considered in the last section would require embedded microphones inside the network as there are no radiative loss channels to the surrounding free space. We circumvent this problem by opening the system of pipes at each vertex thus connecting the underlying structure to the surface through a cavity as shown in Fig.~\ref{fig:exp}(a,c) for the isotropic and anisotropic configuration, respectively. The open cavities at the interface provide experimental access to probe the acoustic waveguide modes inside the network; we measure the diffractive near field which couples resulting in a localised surface wave. This surface mode is imbued with the dispersive characteristics of the underlying graph topology that we detect using scanning microphone techniques.

Acoustic characterisations are performed by measuring the evolution of an acoustic pulse along the surface of the samples. We construct an isotropic and anisotropic sample in accordance with the models discussed in the last section. The samples are manufactured by CNC milling/drilling of several aluminium plates, creating the acoustic path of the fluid (air), see Fig.~\ref{fig:exp}(a,c); For the frequencies/intensities considered the plates act as sound hard (Neumann) boundaries \cite{Beadle2019,Moore2024}. The plates have grooves of square cross section and cylindrical holes that allow the meandering graph topology to be achieved without altering the unit cell length. To obtain the dispersive properties (dispersion curves and isofrequency contours), samples are excited by a tweeter (TFD Near-Field Loudspeaker) mounted adjacent to the side holes of the sample. The loudspeaker is driven by an arbitrary waveform generator (Keysight 33500B), producing single-cycle Sine-Gaussian pulses centred at $f_\text{c} = 5$ kHz, and a broadband amplifier (Cambridge Topaz AM5).

\begin{figure*}
    \centering
    \includegraphics[width = 0.9\textwidth]{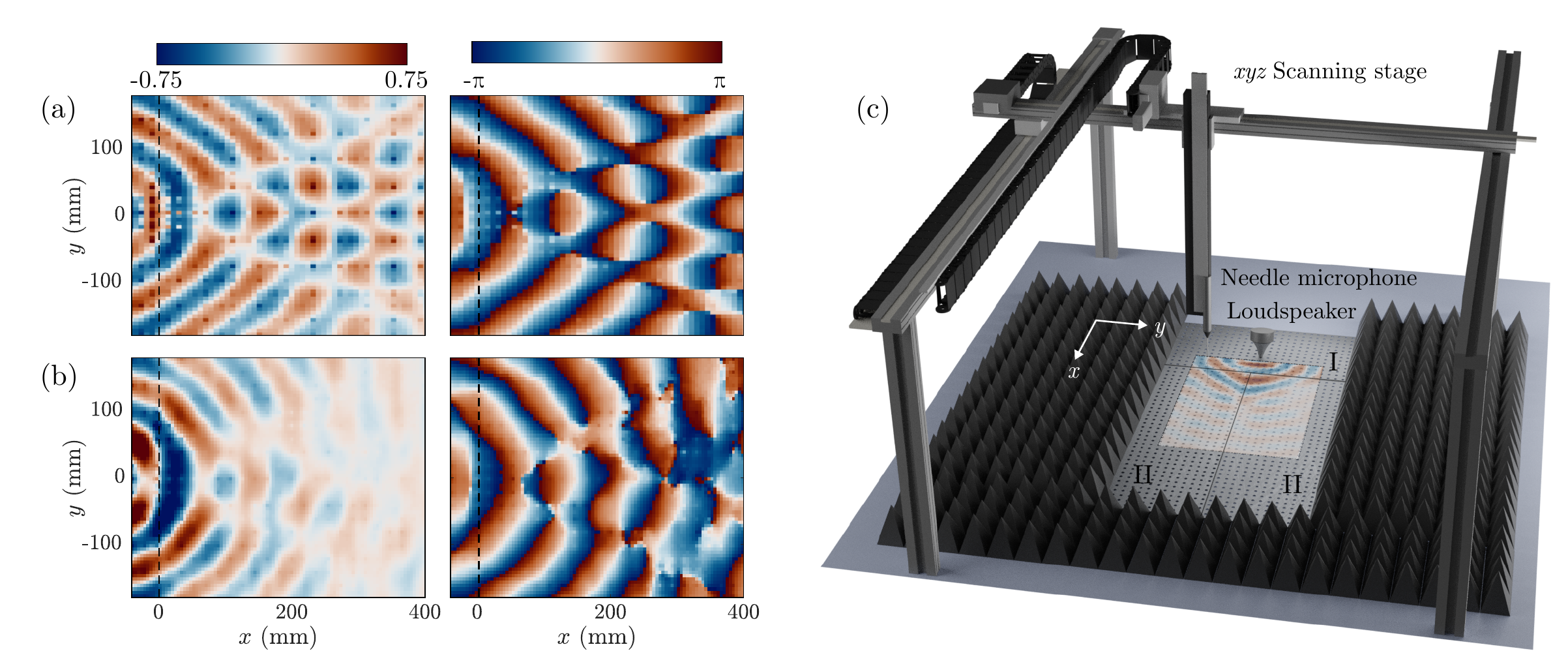}
    \caption{Comparisons of simulation and experiment. (a) Frequency domain simulation with point source excitation and PML boundaries with normalised real pressure field (left) and phase (right) shown at $4.2$ kHz. (b) Experimental results (temporal Fourier transform) of normalised real pressure field (left) and phase (right) $4.2$ kHz. (c) Schematic of scanning stage and plate systems (I) and (II) that are aligned so the holes are in the same plane. Example spatial scan at $4.6$ kHz is shown to highlight the scan area.}
    \label{fig:SimExp1}
\end{figure*}

The acoustic pressure field is measured with a small aperture microphone (Br\"{u}el {\&} Kjær Probe Type 4182 near-field microphone, with a preconditioning amplifier) positioned approximately 1 mm above the sample(s). Acoustic data are recorded by an oscilloscope (Picoscope 5000a) at sampling frequency $f_s = 312.5$ kHz. The microphone is mounted on a motorised \textit{xyz} scanning stage (in-house with Aerotech controllers, a schematic of which is shown in Fig.~\ref{fig:SimExp1}(c)), to spatially map the acoustic signal along an $xy$ area of $34 \times 25$ unit cells, with 3 points per unit cell step-size.  An average was taken over 20 measurements at each spatial position to improve the signal-to-noise ratio. 

Acoustic data are analysed using Fourier techniques; the fast-Fourier Transform (FFT, operator $\mathcal{F}$) of the measured signal, the voltage $V(\boldsymbol{x},t)$, returns the real Fourier amplitude in terms of the wavenumber $\boldsymbol{k} = (k_{x},k_{y})$ and frequency $f$, that is, we perform $\mathcal{F}_{\bf x}(|\mathcal{F}_t(V(\boldsymbol{x},t))|)$.  Time-gated windowing is used to exclude reflected signals; zero-padding by a factor of two is used in the spatial Fourier transform. The resultant experimental Fourier spectra are shown in Fig.~\ref{fig:exp} (b) and (d) for the isotropic and anisotropic case, respectively. We also add the results from FEM simulations for an equivalent infinite, open 2D lattice (dotted lines). Only FEM solutions on or below the sound cone, that is for $2 \pi f \le c k$ with $c$, the sound velocity in air, are plotted. These solutions are non-radiative and couple strongly into localised surface modes predominantly detected in our experimental set-up. In Fig.~\ref{fig:exp} (e) and (g), we show the numerical and experimental isofrequency contours at 4.2 kHz (dashed line in Fig.~\ref{fig:exp} (b,d)) for the isotropic configuration; the spectra closely follow the analytical result of the graph model, Eq.\ \eqref{Graph Free Space Dissipative Properties}, indicated by the dashed line (the sound cone is plotted here as a dotted line). In Fig.\ \ref{fig:exp} (f) and (h), we show the results for the anisotropic configuration; the isofrequency contours exhibit a hyperbolic shape, again matching the quantum graph results, \eqref{Graph Dissipative Properties} outside the sound cone. Solutions inside the sound cone are suppressed - especially in the experimental case - as they are not easily detectable in our near-field set-up. Note that the phase and group velocities, $v_p$ and $v_{g}$, are not co-linear in the anisotropic case (blue and green arrows in Fig.~\ref{fig:exp} (e) and (f)). 

Notable differences in the dispersion relations between the open and closed system (for both isotropic and anisotropic geometries) exist. Firstly, there is no sound-line in the closed system, as it cannot couple to free-space. Secondly, accidental degeneracies (band crossings) have been lifted by connecting to the continuous free space above the metasurface; there is no longer reflection symmetry about the plane of the graph network \cite{chaplain2020delineating}, and the resonance of the pipes, mediated by the diffractive coupling present between the holes on the surface, results in dispersion curves bound by the spoof-plasmon-like frequency \cite{ward2017manipulation}. Despite these differences, that again rise as the graph model is kept as simplistic as possible, the behaviour of negative refraction can be inferred from the QGT, given the predictions of hyperbolic dispersion curves in both systems. The graph model is particularly advantageous here as obtaining the isofrequency contours (Fig.~\ref{fig:SimExp1}(e,f,g,h)) is difficult using finite elements, and only having the dispersion curves in the irreducible Brillouin Zone has well-known drawbacks \cite{craster2012dangers}.


\subsection{Engineering an Interface: Negative Refraction}
Armed with the isofrequency contours (experimental, analytical, and numerical), it is possible to design refraction at an interface between two media. The direction of energy propagation in each media is along $v_{g}$ normal to the isofrequency contours towards the positive gradient as indicated by the green vectors in Fig.\ \ref{fig:exp} (e) and (f). Refraction is then determined in the conventional way through conservation of the tangential component of the wavevector to the interface (Snell's law).  In this section, we demonstrate negative refraction of acoustic modes at the interface between the isotropic and anisotropic media, labelled I and II in the following, see also \cite{lawrie2022quantum} for equivalent results in the graph model. The set-up is shown in Fig.~\ref{fig:SimExp1}. The loudspeaker is mounted above a hole in sample I and the $xy$ scanning area is $24 \times 18$ unit cells, with the source centred 5 unit cells away from the interface. Pyramidal absorbing foam is included to reduce reflections. The anisotropic sample is formed of two plates, as shown in Fig.~\ref{fig:SimExp1}. We perform a Fourier analysis as outlined above and show a comparison between a frequency domain FEM simulation (the `numerical experiment' with perfectly matched layer (PML) boundaries and a monopole point source) and the spatial response at fixed frequency observed in the experiment. We show both the real pressure amplitude and phase of the solutions from FEM simulations and the experiment in Fig.~\ref{fig:SimExp1}(a,b), respectively. Negative refraction manifests through the interference of the wavefronts from the point source, highlighted by the discontinuities in phase.  The resulting wave patterns are similar to other hyperbolic metamaterials \cite{shen2015broadband,yves2023twist}. We note that negative refraction can only be observed indirectly here through the interference of the wavefronts over the surface. 

\begin{figure}
    \centering
    \includegraphics[width=0.45\textwidth]{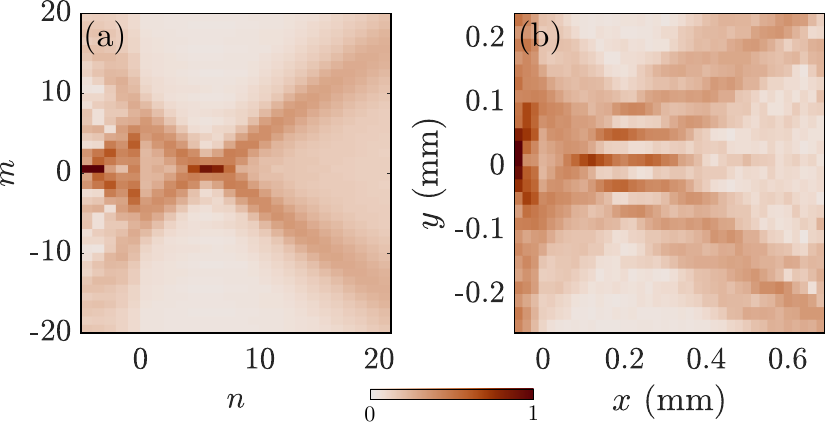}
    \caption{Comparison of graph model and FEM. (a) Absolute field from closed graph model at vertices corresponding to plots in Fig.~\ref{fig:SimExp1}. (b) Normalised absolute pressure field from FEM solution extracted at equivalent vertex positions (intersections of square paths). }
    \label{fig:graphFem}
\end{figure}

To make this effect more obvious, we show in Fig.~\ref{fig:graphFem}(a) the absolute field at each vertex of the network calculated using the (closed) graph model. We compare this in Fig.\ \ref{fig:graphFem}(b) with the equivalent absolute pressure field extracted from the FEM simulation for the open model at the intersection points of the pipes. These positions are not accessible with the microphone in the experiment. Good comparison of the FEM results with the experiment at the surface (Fig.\ \ref{fig:SimExp1}) as well as with the graph model for points inside the network (Fig.\ \ref{fig:graphFem}) clearly demonstrate negative refraction and the utility of the graph model. 


\section{Conclusion}

We have successfully applied quantum graph theory to the design of acoustic metamaterials as a network of interconnected space-coiled waveguides. We purposefully retain the generality of the graph model highlighting its usefulness for all scalar wave regimes that can be reduced to systems of coupled 1D propagation problems. The speed, accuracy, and flexibility of the model allows large parameter spaces to be quickly investigated, providing insight into the dispersive properties of mesh-like metamaterials. This was experimentally verified by extending the model geometry to include an open system by way of connecting the structure to an acoustic metasurface of coupled resonant cavities, characterising two classes of structures (isotropic and anisotropic). We demonstrate negative refraction at an interface without explicit negative index or resonant behaviour. The design of the experiment resulting in a hyperbolic dispersion of the anisotropic configuration was entirely motivated by similar phenomena observed in the closed graph \cite{lawrie2022quantum}. 
This establishes acoustic pipe networks together with quantum graph modelling as a new paradigm towards metamaterial design and construction.

\begin{acknowledgements} 
\noindent The authors wish to acknowledge financial support from the Engineering and Physical Sciences Research Council (EPSRC) via PhD funding for T.M.L. and via the EPSRC Centre for Doctoral Training in Metamaterials (Grant No. EP/ L015331/1). D.B.M. and T.A.S. acknowledge the financial support of Defence Science and Technology Laboratory (Dstl) through grants No. DSTLXR1000154754 and No. AGR 0117701. G.J.C. gratefully acknowledges financial support from the Royal Commission for the Exhibition of 1851 in the form of a Research Fellowship. The authors would like to thank the Isaac Newton Institute for Mathematical Sciences, Cambridge, for support and hospitality during the program {\em Multiple Wave Scattering} supported by EPSRC grant no EP/R014604/1. All data created during this research are available upon reasonable request to the corresponding author. ‘For the purpose of open access, the author has applied a ‘Creative Commons Attribution (CC BY) licence to any Author Accepted Manuscript version arising from this submission’.
\end{acknowledgements}

















\end{document}